\documentclass[twocolumn]{aastex631}
\usepackage{graphicx, color} 
\begin{document}

\title{An Investigation of New Brown Dwarf Spectral Binary Candidates From the Backyard Worlds: Planet 9 Citizen Science Initiative}

\author[0009-0002-3936-8059]{Alexia Bravo}
\affil{United States Naval Observatory, Flagstaff Station, 10391 West Naval Observatory Rd., Flagstaff, AZ 86005, USA}

\author[0000-0002-6294-5937]{Adam C.~Schneider}
\affil{United States Naval Observatory, Flagstaff Station, 10391 West Naval Observatory Rd., Flagstaff, AZ 86005, USA}

\author[0000-0001-8170-7072]{Daniella Bardalez Gagliuffi}
\affil{Department of Physics \& Astronomy, Amherst College, 25 East Drive, Amherst, MA 01003, USA}

\author[0000-0002-6523-9536]{Adam J. Burgasser}
\affil{Department of Astronomy \& Astrophysics, University of California San Diego, La Jolla, CA 92093, USA}

\author[0000-0002-1125-7384]{Aaron M. Meisner}
\affil{NSF's National Optical-Infrared Astronomy Research Laboratory, 950 N. Cherry Ave., Tucson, AZ 85719, USA}

\author[0000-0003-4269-260X]{J. Davy Kirkpatrick}
\affil{IPAC, Mail Code 100-22, Caltech, 1200 E. California Blvd., Pasadena, CA 91125, USA}

\author[0000-0001-6251-0573]{Jacqueline K. Faherty}
\affil{Department of Astrophysics, American Museum of Natural History, Central Park West at 79th St., New York, NY 10024, USA}

\author[0000-0002-2387-5489]{Marc J. Kuchner}
\affil{Exoplanets and Stellar Astrophysics Laboratory, NASA Goddard Space Flight Center, 8800 Greenbelt Road, Greenbelt, MD 20771, USA}

\author[0000-0001-7896-5791]{Dan Caselden}
\affil{Department of Astrophysics, American Museum of Natural History, Central Park West at 79th St., New York, NY 10024, USA}

\author[0000-0003-4864-5484]{Arttu Sainio}
\affil{Backyard Worlds: Planet 9, USA}

\author[0000-0002-7389-2092]{Les Hamlet}
\affil{Backyard Worlds: Planet 9, USA}

\author{The Backyard Worlds: Planet 9 Collaboration}

\begin{abstract} 
We present three new brown dwarf spectral binary candidates: CWISE J072708.09$-$360729.2, CWISE J103604.84$-$514424.4, and CWISE J134446.62$-$732053.9, discovered by citizen scientists through the Backyard Worlds: Planet 9 project. Follow-up near-infrared spectroscopy shows that each of these objects is poorly fit by a single near-infrared standard. We constructed binary templates and found significantly better fits, with component types of L7+T4 for CWISE J072708.09$-$360729.2, L7+T4 for CWISE J103604.84$-$514424.4, and L7+T7 for CWISE J134446.62$-$732053.9. However, further investigation of available spectroscopic indices for evidence of binarity and large amplitude variability suggests that CWISE J072708.09$-$360729.2 may instead be a strong variability candidate. Our analysis offers tentative evidence and characterization of these peculiar brown dwarf sources, emphasizing their value as promising targets for future high-resolution imaging or photometric variability studies.

\end{abstract}

\section{Introduction} 
Brown dwarfs occupy a unique space between stars and planets, possessing masses below the threshold required for sustained hydrogen fusion in their cores \citep{1962AJ.....67S.579K,1963ApJ...137.1121K,1963PThPh..30..460H}. Their cool temperatures and intrinsic faintness make them difficult to detect, and nearby brown dwarfs are continuing to be discovered (e.g., \citealt{marocco2019, best2020, meisner2020a, meisner2020b, bardalez2020, kirkpatrick2021, kirkpatrick2021b, schneider2021, kota2022, lodieu2022, schapera2022}).  The Backyard Worlds: Planet 9 project \citep{kuchner2017} leverages the collaboration between citizen and professional scientists to locate nearby substellar objects by identifying and analyzing moving objects in images from the Wide-field Infrared Survey Explorer \citep{wright2010, mainzer2014}. Multi-epoch WISE images and motion measurements from WISE data (CatWISE 2020; \citealt{marocco2021}) have allowed for numerous discoveries missed by previous surveys, many of which were based solely on infrared colors or limited to nearby objects with very large proper motions.  In addition to finding new nearby brown dwarfs, the Backyard Worlds project has been adept at discovering unusual substellar objects, such as old, low-metallicity subdwarfs \citep{schneider2020, meisner2021, brooks2022} and co-moving companions \citep{faherty2020, rothermich2021, jalowiczor2021, faherty2021, kiwy2021, softich2022, kiwy2022, gramaize2022}. In this study, we examine three spectrally peculiar brown dwarfs found in the Backyard Worlds program: CWISE J072708.09$-$360729.2 (W0727$-$3607), CWISE J103604.84$-$514424.4 (W1036$-$5144), and CWISE J134446.62$-$732053.9 (W1344$-$7320). These objects were discovered by citizen scientists Dan Caselden, Arttu Sainio, and Les Hamlet. Each were flagged as high-priority for follow-up observations because their photometry and estimated spectral types indicated that they may be nearby ($d <$ 25 pc). Properties of these three sources are provided in Table \ref{tab:props}. 


\begin{deluxetable*}{ccccccc}
\tablecaption{Candidate Spectral Binary Properties\tablenotemark{a}}
\label{tab:props}
\tablehead{\colhead{CatWISE Name} & \colhead{$J$} & \colhead{$K_{\rm S}$} & \colhead{$\mu_{\alpha}$} & \colhead{$\mu_{\delta}$} & \colhead{SpT} & \colhead{SpT} \\ 
\colhead{} & \colhead{(mag)} & \colhead{(mag)} & \colhead{(mas yr$^{-1}$)} & \colhead{(mas yr$^{-1}$)} & \colhead{(single)} & \colhead{(binary)} } 
\startdata
CWISE J072708.09$-$360729.2 & 16.521$\pm$0.017 & 15.344$\pm$0.026 & -54.0$\pm$8.3 & 233.6$\pm$9.3 & T4 & L7+T4 \\
CWISE J103604.84$-$514424.4 & 15.977$\pm$0.012 & 14.776$\pm$0.015 & -108.1$\pm$6.8 & -62.7$\pm$8.0 & T2 & L7+T4 \\
CWISE J134446.62$-$732053.9 & 15.890$\pm$0.087 & 14.185$\pm$0.010 & -34.3$\pm$5.4 & 143.3$\pm$6.7 & T4 & L7+T7 \\
\enddata
\tablenotetext{a}{All near-infrared photometry comes from the Vista Hemisphere Survey (VHS; \citealt{mcmahon2013}), with the exception of the $J$-band magnitude of CWISE J134446.62$-$732053.9, which comes from 2MASS \citep{skrutskie2006}. All proper motions listed are taken from the CatWISE 2020 catalog \citep{marocco2021}.}
\end{deluxetable*}

\section{Observations}
W0727$-$3607, W1036$-$5144, and W1344$-$7320 were observed with the TripleSpec4.1 near-infrared spectrograph \citep{schlawin2014, herter2020} located at the Southern Astrophysical Research (SOAR) telescope. TripleSpec4.1 simultaneously covers 0.8$-$2.4 $\mu$m using six cross-dispersed orders with a resolving power of $\sim$3500. The observations were conducted in AEON queue mode and took place on 16 April 2023 (UT), 10 February 2023 (UT), and 14 March 2023 (UT) for W0727$-$3607, W1036$-$5144, and W1344$-$7320, respectively. The spectra were obtained in an ABBA pattern with 120 second exposures. The total integration times for W0727$-$3607, W1036$-$5144, and W1344$-$7320 were 2880, 2400, and 1920 seconds, respectively. A0 stars were observed immediately after each science target for telluric correction and flux calibration. Data reduction and telluric correction was performed using a modified version of the Spextool data reduction package \citep{vacca2003, cushing2004}. The final reduced spectra are shown in Figure \ref{fig:plot1}.   

\begin{deluxetable*}{lccccccccccc}
\label{tab:standards}
\tablecaption{Properties of Objects Used to Create Binary Templates}
\tablehead{
\colhead{SpT} & \colhead{Disc.} & \colhead{Disc.} & \colhead{Spec.} & \colhead{$J$ } & \colhead{Ref} & \colhead{$\varpi$} & \colhead{Ref} \\
\colhead{  } & \colhead{Name } & \colhead{Ref } & \colhead{Ref } & \colhead{(mag)} & \colhead{  } & \colhead{(mas)  } & \colhead{  } }
\startdata
L0 & 2MASP J0345432$+$254023   & 1 & 15 & 13.924$\pm$0.003 & 23 & 37.89$\pm$0.26 & 27 \\
L1 & 2MASS J21304464$-$0845205  & 2 & 16 & 14.059$\pm$0.002 & 24 & 37.47$\pm$0.31 & 27 \\
L2 & 2MASS J04082905$-$1450334\tablenotemark{a} & 3 & 16 & 14.128$\pm$0.002 & 24 & 45.57$\pm$0.27 & 27 \\
L3 & 2MASSW J1506544$+$132106   & 4 & 17 & 13.211$\pm$0.002 & 25 & 85.43$\pm$0.19 & 27 \\
L4 & 2MASS J21580457$-$1550098  & 2 & 16 & 14.794$\pm$0.004 & 24 & 43.11$\pm$0.91 & 27 \\
L5 & 2MASS J21373742$+$0808463\tablenotemark{b}  & 5 & 18 & 14.644$\pm$0.005 & 25 & 66.37$\pm$0.66 & 27 \\
L6 & 2MASSI J1010148$-$040649   & 6 & 19 & 15.372$\pm$0.005 & 24 & 57.7$\pm$3.6 & 28 \\
L7 & 2MASSI J0825196$+$211552   & 7 & 18 & 15.014$\pm$0.005 & 25 & 93.19$\pm$0.59 & 29 \\
L8 & 2MASSW J1632291$+$190441   & 8 & 17 & 15.823$\pm$0.010 & 25 & 66.29$\pm$1.61 & 29 \\
L9 & DENIS J025503.3$-$470049   & 9 & 20 & 13.122$\pm$0.001 & 24 & 205.43$\pm$0.19 & 27 \\
T0 & WISEPA J015010.86$+$382724.3\tablenotemark{c} & 10 & 10 & 15.901$\pm$0.010 & 25 & 44.6$\pm$3.2 & 28 \\
T1 & SDSS J083717.21$-$000018.0  & 11 & 20 & 16.929$\pm$0.007 & 26 & 29.8$\pm$2.7 & 30 \\
T2 & SDSS J125453.90$-$012247.5  & 11 & 21 & 14.694$\pm$0.002 & 26 & 78.34$\pm$1.07 & 29 \\
T3 & WISEPC J223937.55$+$161716.2\tablenotemark{d} & 10 & 10 & 15.995$\pm$0.010 & 25 & 42.9$\pm$3.0 & 30 \\
T4 & 2MASSI J2254188$+$312349   & 12 & 21 & 15.000$\pm$0.005 & 25 & 72.0$\pm$3.0 & 31 \\
T5 & 2MASS J15031961$+$2525196  & 13 & 21 & 13.621$\pm$0.003 & 25 & 155.78$\pm$0.76 & 27 \\
T6 & SDSSp J162414.37$+$002915.6 & 14 & 22 & 15.187$\pm$0.006 & 25 & 90.9$\pm$1.2 & 32 \\
T7 & 2MASSI J0727182$+$171001   & 12 & 22 & 15.210$\pm$0.006 & 25 & 112.5$\pm$0.9 & 33 \\
T8 & 2MASSI J0415195$-$093506   & 12 & 21 & 15.327$\pm$0.004 & 24 & 175.2$\pm$1.7 & 33 \\
\enddata
\tablenotetext{a}{This object replaces the L2 near-infrared standard (Kelu-1), which is a resolved binary \citep{liu2005}.}
\tablenotetext{b}{This object replaces the L5 near-infrared standard (2MASS J08350622$+$1953050), which has no measured parallax.}
\tablenotetext{c}{This object replaces the T0 near-infrared standard (2MASS J12074717$+$0244249), which is a suspected spectral binary \citep{burgasser2010}.}
\tablenotetext{d}{This object replaces the T3 near-infrared standard (2MASS J12095613$-$1004008), which is a resolved binary \citep{liu2010}}
\tablerefs{(1) \cite{kirkpatrick1997}; (2) \cite{kirkpatrick2008}; (3) \cite{wilson2003}; (4) \cite{gizis2000}; (5) \cite{reid2008}; (6) \cite{cruz2003}; (7)\cite{kirkpatrick2000}; (8) \cite{kirkpatrick1999}; (9) \cite{martin1999}; (10) \cite{kirkpatrick2011}; (11) \cite{leggett2000}; (12) \cite{burgasser2002}; (13) \cite{burgasser2003b}; (14) \cite{strauss1999}; (15) \cite{burgasser2006b}; (16) \cite{bardalezgagliuffi2014}; (17) \cite{burgasser2007}; (18) \cite{burgasser2010}; (19) \cite{reid2006}; (20) \cite{burgasser2006}; (21) \cite{burgasser2004}; (22) \cite{burgasser2006c}; (23) \cite{lawrence2007}; (24) \cite{mcmahon2013}; (25) \cite{dye2018}; (26) \cite{edge2013}; (27) \cite{gaia2023}; (28) \cite{kirkpatrick2021}; (29) \cite{dahn2017}; (30) \cite{best2020}; (31) \cite{manjavacas2013}; (32) \cite{tinney2003}; (33) \cite{dupuy2012} }
\end{deluxetable*}

\section{Analysis}
To determine preliminary spectral types, we compared to the near-infrared L and T dwarf standards from \cite{kirkpatrick2010} and \cite{burgasser2006}, with the exception of the L7 standard, where we use 2MASSI J0825196$+$211552 as recommended in \cite{cruz2018}, as the original L7 near-infrared standard appears to have an unusually low surface gravity and age.  The best fitting standards for each object are shown in the left column of Figure \ref{fig:plot1}, and match reasonably well in the $J$-band region, but show significant discrepancies beyond $\sim$1.5 $\mu$m. To ensure that differences between standards and the observed spectra were not due to uncertainties incurred when stitching orders or some other observational systematic, we calculated synthetic photometric $J-K_S$ colors using the VISTA filter response curves.  We find synthetic colors consistent with the photometric colors in Table 1 to within 1$\sigma$.  For this reason, we constructed binary templates in an attempt to better fit the observed spectra. 

We utilized the same spectral standards as above, with a few exceptions to account for known distance and multiplicity. The L2 near-infrared standard is a resolved binary \citep{liu2005}, so we instead use the L2 object 2MASS J04082905$-$1450334 \citep{wilson2003, bardalezgagliuffi2014}. The L5 near-infrared standard 2MASS J08350622$+$1953050 has no measured parallax, so we instead use the L5 2MASS J21373742$+$0808463 \citep{reid2008, burgasser2010}. The T0 near-infrared standard is suspected to be a spectral binary \citep{burgasser2010, ashraf2022}, thus we employed the T0 WISEPA J015010.86$+$382724.3 \citep{kirkpatrick2011}. Lastly, the T3 standard 2MASS J12095613$-$1004008 is a resolved binary \citep{liu2010}, and we therefore use the T3 WISEPC J223937.55$+$161716.2 \citep{kirkpatrick2011}. The objects used to create spectral binary templates are summarized in Table \ref{tab:standards}. 
 
To generate spectral binary templates, we acquired near-infrared spectra from the the SpeX Prism Library Analysis Toolkit (SPLAT; \citealt{burgasser2017}). We absolute flux calibrated each spectrum using UKIDSS Hemisphere Survey (UHS; \citealt{dye2018}) or Vista Hemisphere Survey (VHS; \citealt{mcmahon2013}) $J$-band photometry, and measured parallaxes \citep{tinney2003, dupuy2012, manjavacas2013, dahn2017, best2020, kirkpatrick2021, gaia2023}. Finally, we added the spectra together and normalized the result to the $J$-band peak between 1.27--1.29 $\mu$m. We compared the resulting templates to each observed spectrum and found the best fits by calculating $\chi^2_{\nu}$ values following \cite{burgasser2010}.

\subsection{Spectral Index Calculations}
Spectral indices are also used to identify potential objects of interest (e.g., young brown dwarfs, spectral binaries, photometric variables). We follow the methods and calculate the indices defined in \cite{burgasser2002, burgasser2006, burgasser2010} and \cite{bardalezgagliuffi2014} for our observed objects (Table \ref{tab:indices}) and a large comparison sample from the SPLAT library. 
 Uncertainties were determined in a Monte Carlo fashion. We use binary index regions from \cite{burgasser2010} 
'and variability regions from \cite{ashraf2022} in 
the following sections.

\begin{deluxetable*}{lcccc}
\label{tab:indices}
\tablecaption{Spectral Index Values}
\tablenum{2}
\tablehead{\colhead{Spectral} & \multicolumn{3}{c}{Object} & \colhead{Ref.}\tablenotemark{a} \\ 
\cline{2-4}
\colhead{Index} & \colhead{W0727$-$3607} & \colhead{W1036$-$5144} & \colhead{W1344$-$7320} } 
\startdata
H$_2$O$-J$ & 0.392$\pm$0.009 & 0.546$\pm$0.009 & 0.467$\pm$0.017 & 1 \\
CH$_4$$-J$ & 0.542$\pm$0.006 & 0.586$\pm$0.004 & 0.579$\pm$0.010 & 1 \\
$J$-Curve & 4.121$\pm$0.084 & 2.970$\pm$0.042 & 3.286$\pm$0.104 & 2 \\
CH$_4$$-H$ & 0.999$\pm$0.011 & 0.822$\pm$0.005 & 0.858$\pm$0.011 & 1 \\
$H$-Bump & 0.852$\pm$0.012 & 1.120$\pm$0.008 & 1.056$\pm$0.016 & 2 \\
$H$-Dip & 0.518$\pm$0.006 & 0.459$\pm$0.003 & 0.451$\pm$0.006 & 3 \\
CH$_4$$-K$ & 0.607$\pm$0.010 & 0.801$\pm$0.009 & 0.909$\pm$0.013 & 1 \\
$K$-Slope & 0.960$\pm$0.010 & 0.975$\pm$0.008 & 0.973$\pm$0.012 & 4 \\
$K/J$ & 0.453$\pm$0.005 & 0.433$\pm$0.003 & 0.559$\pm$0.007 & 1 \\
H$_2$O$-H$ & 0.414$\pm$0.010 & 0.572$\pm$0.006 & 0.622$\pm$0.013 & 1 \\
\hline
Binarity\tablenotemark{b} & weak & strong & strong \\
Variability\tablenotemark{c} & strong & ... & ... \\
\enddata
\tablenotetext{a}{Given reference corresponds to the work where the index was originally defined.}
\tablenotetext{b}{Based on the index criteria defined in \citet{burgasser2010}.}
\tablenotetext{c}{Based on the index criteria defined in \citet{ashraf2022}.}
\tablerefs{(1) \cite{burgasser2006}; (2) \cite{bardalezgagliuffi2014}; (3) \cite{burgasser2010} \cite{burgasser2002} }
\end{deluxetable*}

\section{Results and Discussion}
The observed discrepancies between observed spectra and existing spectral standards can be attributed to various factors, such as binarity, variability, youth, and metallicity, or a combination of these influences, as discussed below. 

\subsection{Binarity}

\subsubsection{Analysis of Binary Templates}

\begin{figure*}
\epsscale{1.15}
\plotone{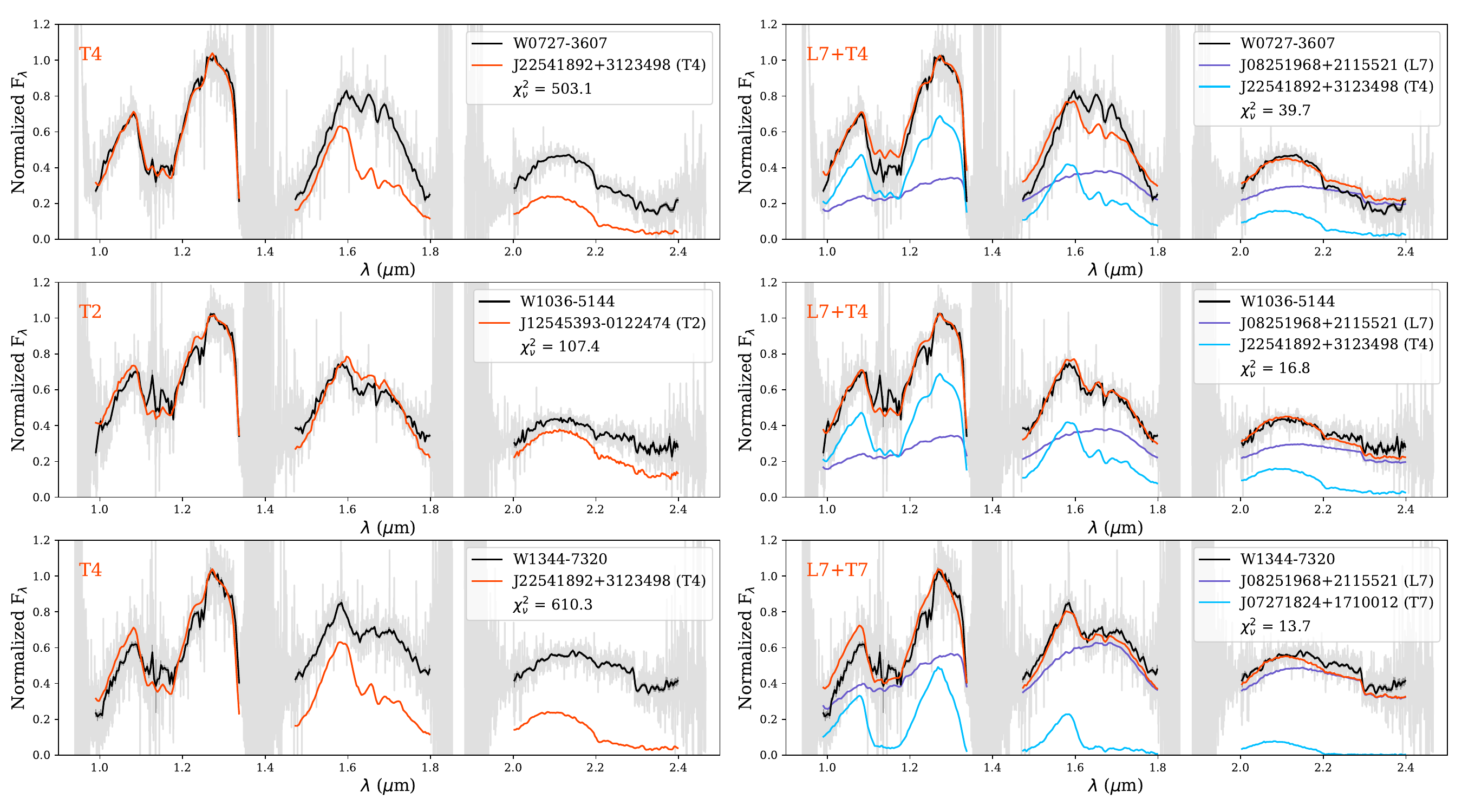}
\caption{ Observed SOAR/TripleSpec4.1 spectra for W0727$-$3607 (top), W1036$-$5144 (middle), and W1344$-$7320 (bottom), with gray lines showing the full resolution and black lines showing the spectra smoothed to the resolution of the standards and binary templates (R$\approx$150) according to their parallax-based absolute fluxes. The left-hand column depicts the best-fitting single standards (orange line) for the spectra of the three candidates while the right-hand column shows both the binary template (orange line) and the standard spectra that compose it (purple line for primary, blue line for secondary).
Reduced $\chi^2_{\nu}$ values are reported for each of the fits.} 
\label{fig:plot1}
\end{figure*}

The right column of Figure \ref{fig:plot1} compares the best-fitting 
binary templates to our sources. 
The discrepancies observed with the single fits beyond approximately 1.5 $\mu$m are greatly reduced with the binary templates, as verified by lower
$\chi^2_{\nu}$ values.
A discussion of the results for each individual object follows.

{\it CWISE J072708.09$-$360729.2}: The best-fitting T4 single standard
exhibits a significantly higher $\chi^2_{\nu}$ = 503 compared to the L7+T4 binary template, which has $\chi^2_{\nu}$ = 39.7. 
Following \cite{burgasser2010}, we evaluate our fits with a one-sided $F$-test, finding $\eta_{\rm SB}$ = 12.7, well above the $\eta_{\rm SB}$ $>$1.34 spectral binary confidence threshold given in that work.
Although the binary template provides a superior fit overall, discrepancies persist, particularly at the $H$-band peak.We retain W0727$-$3607 as a spectral binary candidate, but discuss below that it is also a potential variable source (see Section \ref{sec:conclusions}). 

{\it CWISE J103604.84$-$514424.4}: 
W1036$-$5144 
shows a lower-than-expected peak in the $H$-band and more flux across the $K$-band compared to the T2 standard. These discrepancies are largely resolved in the L7+T4 binary fit,
with only minor deviations in the $J$-band and in the longest wavelengths of the $K$-band. The binary fit
yields a significantly lower $\chi^2_{\nu}$ = 16.8, compared to the $\chi^2_{\nu}$ = 107.4 for the single fit, and
yields $\eta_{\rm SB}$ = 6.4, consistent with the spectral binary criteria specified in \cite{burgasser2010}.

{\it CWISE J134446.62$-$732053.9}: 
The single standard fit to W1344$-$7320
has $H$- and $K$-bands that are underluminous, leading to a relatively poor $\chi^2_{\nu}$ = 610. The L7+T7 binary template 
yields a significantly improved fit with $\chi^2_{\nu}$ = 13.7. Although there is a slight discrepancy in $Y$-band, where the template is too bright compared to the observed spectrum, the remainder of the $J$-, $H$-, and $K$-band regions exhibit a well-matched morphology.
Our $\chi^2_{\nu}$ values correspond to 
$\eta_{\rm SB}$ = 44.5, which again satisfies the spectral binary criteria specified in \cite{burgasser2010}.

\subsubsection{Analysis of Binary Indices}
\label{sec:bindices}

\begin{figure*}
\epsscale{1.15}
\plotone{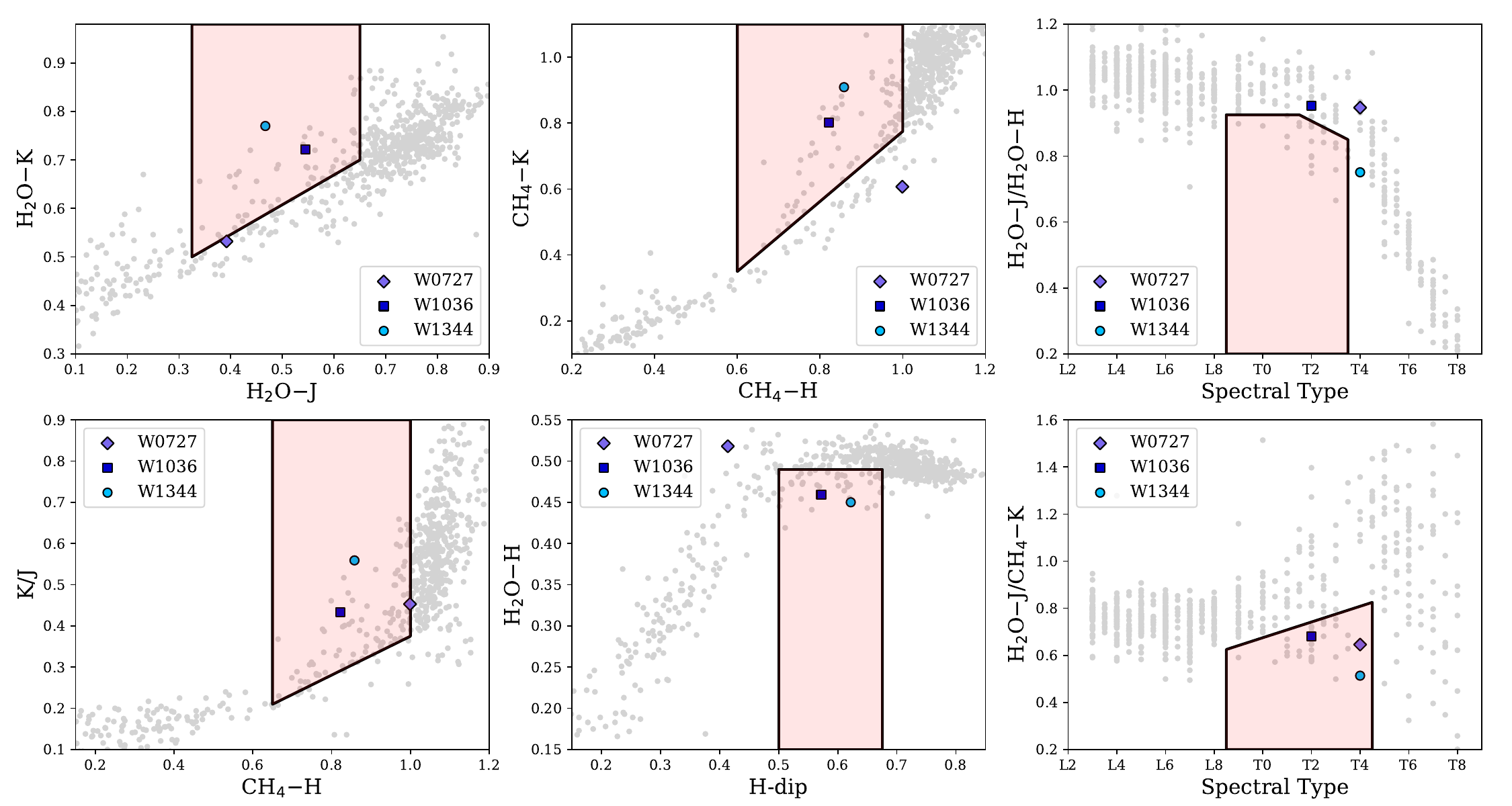}
\caption{Index-index plots highlighting regions defined in \cite{burgasser2010, bardalez2015} as indicative of spectral binarity (red shaded regions). Background grey points are SPLAT objects with spectral types between L3 and T8. The three targets of this study are shown in the legend.} 
\label{fig:binindex}
\end{figure*}

The spectral indices shown to be indicative of binarity in \cite{burgasser2010} are shown in Figure \ref{fig:binindex}.
That study designated sources that satisfied at least three index criteria as ``strong'' candidates, and two index criteria as ``weak'' candidates.

{\it CWISE J072708.09$-$360729.2}: W0727$-$3607 falls within two of five regions indicative of binarity,
and are very close to the edges of these regions.
This classifies the source as a weak candidate.
Note that we exclude for consideration the H$_2$O$-J$/H$_2$O$-H$ versus spectral type comparison as it is not applicable for spectral types later than T3.5.
Along with the mediocre fit of the L7+T4 binary template, this result suggests 
a factor other than binarity is responsible for this object's unusual spectrum.

{\it CWISE J103604.84$-$514424.4}: 
W1036$-$5144 satisfies five of six regions, making it a strong binary candidate.
Considering the well-fit binary template,
binarity is a strong possibility for this source.

{\it CWISE J134446.62$-$732053.9}: W1344$-$7320 
satisfies five of five binary regions (excluding H$_2$O$-J$/H$_2$O$-H$ versus spectral type), making it a strong binary candidate. Again, the 
well-fit binary template of L7+T7 makes
this object a likely spectral binary.

\subsubsection{Distance Estimation}
If these objects are indeed spectral binaries, then updated distance estimates can be obtained by scaling the best-fitting absolutely flux calibrated binary templates to each object's observed photometric magnitudes. Using $J$-band photometry from VHS for W0727$-$3607 and W1036$-$5144 and 2MASS \citep{skrutskie2006} for W1344$-$7320, we find estimated distances of 34$\pm$7 pc, 27$\pm$5 pc, and 20$\pm$4 pc, respectively.

\subsection{Variability}
\begin{figure*}
\epsscale{1.15}
\plotone{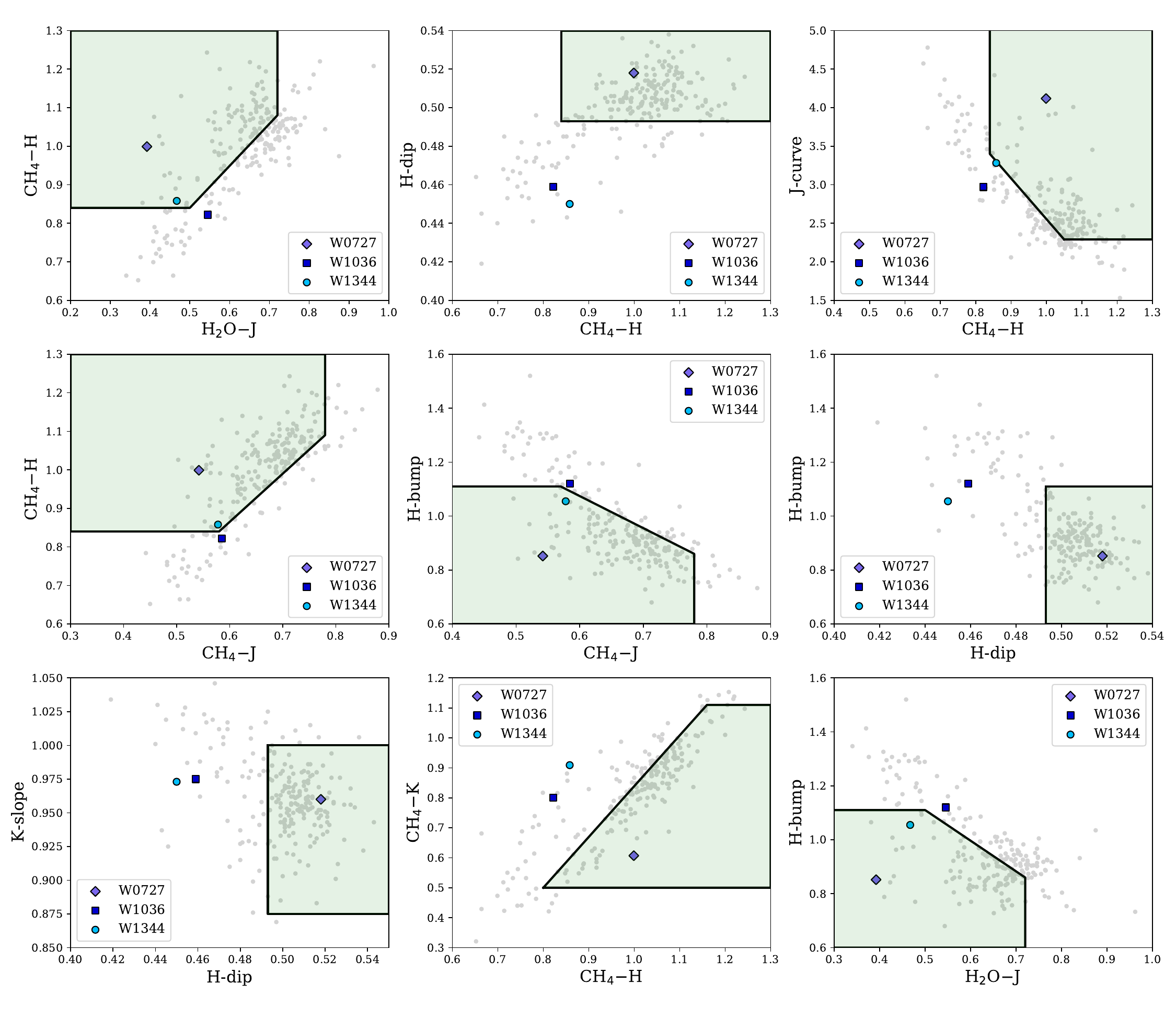}
\caption{Index-index plots highlighting the regions indicative of photometric variability defined in \cite{ashraf2022} (green shaded regions). Background grey points are SPLAT objects with spectral types between L7 and T3, with our three observed targets indicated shown in the legend.} 
\label{fig:varindex}
\end{figure*}

There is evidence that some spectral binary candidates are instead single stars with inhomogeneous atmospheres as indicated via their photometric or spectroscopic variability (e.g., 2MASS J21392676$+$0220226; \citealt{radigan2012,2013AJ....145...71K}). We applied the spectral index criteria introduced in \cite{ashraf2022} to analyze the spectra of W0727$-$3607, W1036$-$5144, and W1344$-$7320 and evaluate their potential as strong variable sources. This method uses single low-resolution spectra to empirically identify spectral indices that may be indicative of variability based on known variable sources. The idea is that cloudy or patchy layers at different temperatures in a brown dwarf atmosphere that lead to photometric variability may have measurable effects in the emergent spectra of these objects. Note that we do not examine any index-index criteria that use the H$_2$O$-$K index, as the region of the numerator range (1.975$-$1.995 $\mu$m) is especially noisy in our TripleSpec4.1 spectra. In Figure \ref{fig:varindex}, there are nine total index-index correlation plots, as discussed below. \cite{ashraf2022} only considered objects that satisfied all of their variability index criteria as ``strong'' candidates, and those that satisfied all but one of their outlined criteria as ``weak'' candidates.

\subsubsection{Analysis of Variable Indices}
\label{sec:vindices}

{\it CWISE J072708.09$-$360729.2}: 
W0727$-$3607 falls within all of the designated regions indicating potential variability.
Coupled with the weak spectral binary designation and less accurate match to binary templates,
variability appear to be at least partly responsible for the unusual shape of this object's spectrum. We investigated potential variability using WISE single-exposure photometry following the methods in \cite{schneider2023}. We found no clear signs of variability in the single-exposure WISE data, with W1 and W2 photometric standard deviation values within 1$\sigma$ of the median value for objects with similar magnitudes. However, we note that with magnitudes of W1 = 14.639$\pm$0.018 mag and W2 = 13.862$\pm$0.15 mag, W0727$-$3607 may be too faint to detect significant variability in WISE data, as no variables were identified in \cite{schneider2023} with magnitudes as faint as these.

{\it CWISE J103604.84$-$514424.4}: W1036$-$5144 consistently lies outside the range indicative of potential variability, supporting the spectral binary hypothesis.

{\it CWISE J134446.62$-$732053.9}: W1344$-$7320 satisfies four of the nine variability regions, below the weak candidate threshold of \cite{ashraf2022}
Again, the spectral binary hypothesis is favored
for this object.

\section{Conclusions}
\label{sec:conclusions}

In this study, we presented three new brown dwarf spectral binary candidates: CWISE J072708.09$-$360729.2, CWISE J103604.84$-$514424.4, and CWISE J134446.62$-$732053.9. By constructing binary templates and comparing them to the observed spectra, we found significantly better fits, revealing component types of T0+T7 for CWISE J072708.09$-$360729.2, L7+T4 for CWISE J103604.84$-$514424.4, and L7+T7 for CWISE J134446.62$-$732053.9. However, our investigation of variability indices suggests that CWISE J072708.09$-$360729.2 is a strong variability candidate. 

The rarity of brown dwarf-brown dwarf binary systems, as consistently highlighted by statistical studies \citep{burgasser2007, radigan2013, abertasturi2014, opitz2016, fontanive2018}, underscores the importance of pinpointing more such systems. Moreover, the ability to identify spectral binaries offers the advantage that their orbital period are sufficiently short to be amenable to dynamical mass follow-up
\citep{2012ApJ...757..110B,bardalez2015,2016ApJ...827...25B,2020MNRAS.495.1136S}. 
Similarly, variable brown dwarfs provide opportunities to study their rotational and atmospheric dynamics, particularly through multi-wavelength studies \citep{2012ApJ...760L..31B,2013ApJ...768..121A,radigan2013}. 
Future work includes confirming the binary and/or variability nature of these objects through high resolution imaging and photometry, astrometric, and radial velocity monitoring. 
Increasing the sample of close binary and variable brown dwarfs has the potential to advance our understanding of substellar formation, evolution, and dynamics.

\begin{acknowledgements}
The Backyard Worlds: Planet 9 team would like to thank the many Zooniverse volunteers who have participated in this project. We would also like to thank the Zooniverse web development team for their work creating and maintaining the Zooniverse platform and the Project Builder tools. This work is based on observations obtained at the Southern Astrophysical Research (SOAR) telescope, which is a joint project of the Minist\'{e}rio da Ci\^{e}ncia, Tecnologia e Inova\c{c}\~{o}es (MCTI/LNA) do Brasil, the US National Science Foundation’s NOIRLab, the University of North Carolina at Chapel Hill (UNC), and Michigan State University (MSU). This material is based upon work supported by the National Science Foundation under Grant No. 2007068, 2009136, and 2009177. This publication makes use of data products from the Wide-field Infrared Survey Explorer, which is a joint project of the University of California, Los Angeles, and the Jet Propulsion Laboratory/California Institute of Technology, funded by the National Aeronautics and Space Administration. This publication also makes use of data products from NEOWISE, which is a project of the Jet Propulsion Laboratory/California Institute of Technology, funded by the Planetary Science Division of the National Aeronautics and Space Administration.
\end{acknowledgements}

\software{SPLAT \citep{burgasser2017}, Spextool \citep{vacca2003,cushing2004}}
\facilities{SOAR, WISE}

\bibliography{references}{}
\bibliographystyle{aasjournal}

\end{document}